\documentclass[english,aps,manuscript]{revtex4}
\usepackage[latin1]{inputenc}
\usepackage{amssymb}

\makeatletter

\newcommand{\bee}{\begin{equation}}
\newcommand{\ee}{\end{equation}}
\newcommand{\beea}{\begin{eqnarray}}
\newcommand{\eea}{\end{eqnarray}}

\usepackage{babel}
\makeatother
\begin{document}
\textbf{\Large On Integrating Out Heavy Fields in SUSY Theories} 

\vspace{0.3cm}

\begin{center}{\large S. P. de Alwis$^{\dagger}$ }\end{center}{\large \par}

\begin{center}Physics Department, University of Colorado, \\
 Boulder, CO 80309 USA\end{center}

\begin{center}\vspace{0.3cm}\end{center}

\begin{center}\textbf{Abstract}\end{center}

We examine the procedure for integrating out heavy fields in supersymmetric
(both global and local) theories. We find that the usual conditions
need to be modified in general and we discuss the restrictions under
which they are valid. These issues are relevant for recent work in
string compactification with fluxes.

\vspace{0.3cm}


\vfill

$^{\dagger}$ {\small e-mail: dealwis@pizero.colorado.edu}{\small \par}

\eject

~

In a theory containing both heavy (mass $M$) and light (mass $m\ll M$)
fields, one may derive an effective field theory \cite{Weinberg:1980wa}
valid for energy scales $E\ll M$, by doing the functional integral
over the heavy fields (as well as light field modes with frequencies
greater than $M$). If one is just considering the classical approximation
then this just means solving the classical equation for the heavy
field in terms of the light field and substituting back into the action.
So if the potential for the heavy ($\Phi$) and light ($\phi$) fields
is $V(\Phi,\phi)=\frac{1}{2}M^{2}\Phi^{2}+\tilde{V}(\Phi,\phi)$ the
equation of motion for the heavy field gives,\[
\Phi=\frac{1}{\square-M^{2}}\frac{\partial\tilde{V}}{\partial\Phi}=-\frac{1}{M^{2}}\frac{\partial\tilde{V}}{\partial\Phi}+\frac{1}{M^{2}}O(\frac{\square}{M^{2}}\frac{\partial\tilde{V}}{\partial\Phi}).\]
 In other words up to terms in the derivative expansion that maybe
ignored at energy $E\ll M$, $\Phi$ is a solution of the equation\begin{equation}
\frac{\partial V}{\partial\Phi}=0\label{nonsusy}\end{equation}

Consider now a globally supersymmetric theory of chiral scalar fields
with superspace action \begin{equation}
S=\int d^{4}xd^{4}\theta\bar{\Phi^{i}}\Phi^{i}+(\int d^{4}xd^{2}\theta W(\Phi^{i})+c.c.].\label{rigidsusy}\end{equation}

The superfield equations of motion are (with $D^{2}=D^{\alpha}D_{\alpha}$where
$D_{\alpha}$ is the spinor covariant derivative)%
\footnote{We use the conventions of \cite{Wess:1992cp}. For the derivations
below see also \cite{Gates:1983nr}.%
}, \begin{equation}
-\frac{1}{4}\bar{D}^{2}\bar{\Phi^{i}}+\frac{\partial W}{\partial\Phi^{i}}=0.\label{rigidsusyeqn}\end{equation}

The component form of this is obtained by successive application of
the spinorial covariant derivative and setting $\theta=\bar{\theta}=0$
(indicated below by the symbol $|$). In particular the auxiliary
field equation (from $D^{0}$) and the scalar field (from $-\frac{1}{4}D^{2}$)
become (after setting the fermions to zero and using $D^{2}\bar{D^{2}}\bar{\Phi}=16\square\bar{\Phi}$)\begin{eqnarray}
\bar{F^{i}}+\frac{\partial W}{\partial\Phi^{i}}| & =0\label{susyeqn1}\\
\square\bar{\Phi^{i}|}+\sum_{j}\frac{\partial W}{\partial\Phi^{i}\partial\Phi^{j}}|F^{j} & =0\label{susyeqn2}\end{eqnarray}

Now suppose that we wish integrate out a heavy field with say $i=H$
to get an effective theory for the light fields with $i=l$. For example
one might have $W=\frac{1}{2}M\Phi^{H2}+\frac{1}{2}\lambda_{ll'}\Phi^{H}\Phi_{l}\Phi_{l'}+W_{L}(\Phi_{l})$.
Then the analog of (\ref{nonsusy}) (see for example \cite{Intriligator:1995au})
is to require that \begin{equation}
\frac{\partial W}{\partial\Phi^{H}}=0.\label{susyextrem}\end{equation}
Taking components of this equation, (and ignoring the fermionic terms)\begin{eqnarray}
\frac{\partial W}{\partial\Phi^{H}}| & =0\label{extremum1}\\
\sum_{j}\frac{\partial^{2}W}{\partial\Phi^{H}\partial\Phi^{l}}|D^{2}\Phi^{l}+\frac{\partial^{2}W}{\partial\Phi^{H}\partial\Phi^{H}}|\frac{\partial\bar{W}}{\partial\bar{\Phi}^{H}}| & =0\label{extremum2}\\
\nonumber \end{eqnarray}
where in the second (obtained by acting with $D^{2}$ and setting
fermions to zero) we have used (\ref{susyeqn1}) for $i=H$. The second
condition is of course the requirement that the potential $V=\sum_{i}F^{i}\bar{F^{i}}$
be extremized with respect to $\Phi_{H},$ which is what one would
impose in a non-supersymmetric theory (see (\ref{nonsusy})). However
here we have the additional condition (\ref{extremum1}) which in
conjunction with (\ref{extremum2}) and the equatinon of motion for
the light field leads to \[
\sum_{j}\frac{\partial^{2}W}{\partial\Phi^{H}\partial\Phi^{l}}|\frac{\partial W}{\partial\Phi^{l}}|=0\]

But (after solving for the heavy field using (\ref{susyextrem}) this
is a constraint on the light fields in theories where there is a non-zero
coupling between the heavy and light fields as in the above example.
In particular if there is only one light field it imposes the condition
$\frac{\partial W}{\partial\Phi^{l}}|=0$, meaning that the light
scalar is also at the minimum of the potential. In other words one
does not get an effective potential for the light field - the only
consistent result of integrating out the heavy field is that all fields
are sitting at the SUSY minimum (if it exists). When there is more
than one light field this is not necessarily the case but nevertheless
the light field space is constrained. If we had kept the fermionic
terms then this constraint would be a relation between the bosonic
components and squares of fermionic components of the light fields.

To see where this comes from let us write the superpotential for the
heavy light theory as\begin{equation}
W(\Phi^{H},\Phi^{l})=\frac{1}{2}M\Phi^{H2}+\tilde{W}(\Phi^{H},\Phi^{l}).\label{model}\end{equation}

Operating on the (conjugate of the) equation of motion (\ref{rigidsusyeqn})
with ($-\frac{1}{4}D^{2}$), using (\ref{rigidsusyeqn}) again and
rearranging we have,\[
\Phi^{H}=\frac{1}{\square-M^{2}}(M\frac{\partial\tilde{W}}{\partial\Phi^{H}}+\frac{\bar{D}^{2}}{4}\frac{\partial\bar{\tilde{W}}}{\partial\bar{\Phi}^{H}})\]
 Expanding the inverse Klein-Gordon operator as before we can rewrite
this as,\[
M\Phi^{H}+\frac{\partial\tilde{W}}{\partial\Phi^{H}}=-\frac{\bar{D}^{2}}{4M}\frac{\partial\bar{\tilde{W}}}{\partial\bar{\Phi}^{H}}+O(\frac{\square}{M^{2}}(...)).\]

So to the lowest order in the space-time derivative (momentum) expansion
what we get for the equation determining the heavy field in terms
of the light is,\begin{equation}
\frac{\partial W}{\partial\Phi^{H}}=-\frac{\bar{D}^{2}}{4M}\frac{\partial\bar{\tilde{W}}}{\partial\bar{\Phi}^{H}},\label{susyheavy}\end{equation}

rather than (\ref{susyextrem}). To get the latter one needs the additional
assumption that the possible values of $\bar{D^{2}}\frac{\partial\bar{\tilde{W}}}{\partial\bar{\Phi}^{H}}$
are small compared to $M$. For instance in the above example (in
the paragraph after (\ref{susyeqn2}) this means that we need $\Phi_{l}<<M$.
This of course is what one would expect. However as we pointed out
earlier in this same approximation the light field space appears to
be constrained. 

Let us be even more specific and consider the model (\ref{model})
with $\tilde{W}=\frac{1}{2}HL^{2}$. (We've relabelled the heavy field
as $H$ and the light field as $L$.) Then (\ref{susyheavy}) becomes,\begin{eqnarray}
\frac{\partial W}{\partial H} & =MH+\frac{1}{2}L^{2} & =-\frac{\bar{D}^{2}}{4M}(\frac{1}{2}\bar{L}^{2})\nonumber \\
 & =-\frac{1}{4M}(\bar{D}^{\dot{\alpha}}\bar{\bar{L}D_{\dot{\alpha}}}\bar{L} & +\bar{L}\bar{D}^{2}\bar{L)}\label{model1}\end{eqnarray}
 Again we see that the strict imposition of $\partial W/\partial H=0$
leads to the constraint on the light field space that we found above.
To see in what approximation this equation is valid let us solve it
for $H$ (giving $H=-L^{2}/M$) and then plug it back into the RHS
of the last equality of (\ref{model1}) after using the light field
equation. The bosonic term is then $L^{2}\frac{|L|^{2}}{M^{2}}$and
is small when $|L|<<M$. 

It is perhaps worthwhile looking at this example in component form.
The superspace Lagrangian in the above example is \begin{equation}
\int d^{4}\theta(\bar{H}H+\bar{L}L)+\left[\int d^{2}\theta\frac{1}{2}(MH^{2}+HL^{2})+c.c.\right]\label{example}\end{equation}

In components one has, writing the scalar and F components of $H=(A,$F)
and of $L=(a,f)$ and ignoring the fermion terms,\begin{eqnarray}
L & = & \bar{A}\square A+\bar{a}\square a+\bar{F}F+\bar{f}f+\frac{1}{2}(Fa^{2}+\bar{F}\bar{a}^{2})+(Aaf+\bar{A}\bar{a}\bar{f})\nonumber \\
 & + & M(AF+\bar{A}\bar{F})\label{complag}\end{eqnarray}

The heavy field equations are \begin{eqnarray}
-\square\bar{A} & = & MF+af\label{Abar}\\
\bar{F} & = & -\frac{a^{2}}{2}-MA\label{Fbar}\end{eqnarray}

Solving them we get after expanding in powers of $\square/M^{2}$,

\begin{eqnarray*}
\bar{A} & = & \frac{1}{2M^{2}}(-M\bar{a}^{2}+2af)+\frac{1}{M^{2}}O(\frac{\square}{M^{2}})\\
\bar{F} & = & -\frac{\bar{a}\bar{f}}{M}+\frac{1}{M^{2}}O(\frac{\square}{M^{2}})\end{eqnarray*}

It is easily checked that these are precisely the equations that would
be obtained by looking at the components of the superfield equation
(\ref{susyheavy}). Plugging these equations into (\ref{complag})
we get the light field potential,\[
-V=\bar{f}f(1+\frac{\bar{a}a}{M^{2}})-\frac{1}{2M}(fa^{3}+\bar{f}\bar{a}^{3})\]

Eliminating the light auxiliary field we have\[
V=\frac{|a|^{6}}{4M^{2}}(1+\frac{|a|^{2}}{M^{2}})^{-1}\]

If we had just imposed the usual condition $\partial W/\partial H=0$
we would not have got the $|a|^{2}/M^{2}$term in the parenthesis.
This means that this condition gives the correct result for the potential
only for small values of the field $|a|<<M$.

To derive the analog of (\ref{susyheavy}) in supergravity we use
the formalism in chapter 8 of \cite{Gates:1983nr} (though we remain
with the conventions of \cite{Wess:1992cp}). In this formalism the
matter equations can be derived by replacing the supervielbein determinant
by the so-called chiral compensator field $\phi$ and treating the
coupling of the matter fields as in flat superspace. Thus the supercovariant
derivative $D_{\alpha}$is the flat space one and satisfies $D^{3}=0$.
Also acting on chiral fields we have $(D^{2}\bar{D}^{2}/16)\Phi=\square\Phi$.
The action is (with $M_{p}^{2}=1$)\begin{equation}
S=-3\int d^{4}xd^{4}\theta\bar{\phi}\phi e^{-K/3}+(\int d^{4}xd^{2}\theta\phi^{3}W+h.c.)\label{sugraaction}\end{equation}

where the superpotential is a holomorphic function of the chiral scalar
fields $W=W(\chi^{i})$ and the Kaehler potential is a real function
$K=K(\chi,\bar{\chi})$. From this action one obtains the following
equations of motion.\begin{eqnarray*}
\frac{1}{4}\bar{D^{2}}\bar{\chi}^{\bar{k}}+\frac{1}{4}K^{\bar{k}i}(K_{i\bar{j}l}-\frac{2}{3}K_{i\bar{j}}K_{\bar{l}})\bar{D}^{\dot{\alpha}}\bar{\chi}^{\bar{j}}\bar{D}_{\dot{\alpha}}\bar{\chi}^{\bar{l}} & +\frac{1}{2}\frac{\bar{D}^{\dot{\alpha}}\bar{\phi}}{\bar{\phi}}\bar{D}_{\dot{\alpha}}\bar{\chi}^{\bar{k}}\\
=e^{K/3}\frac{\phi^{2}}{\bar{\phi}}K^{\bar{k}i}D_{i}W\\
\frac{1}{4}\bar{D^{2}}(\bar{\phi}e^{-K/3})=-\phi^{2}W(\chi)\end{eqnarray*}

Using the identity above (\ref{sugraaction}) we get,\begin{equation}
\square\bar{\chi}^{\bar{k}}=-\frac{D^{2}}{16}[K^{\bar{k}i}(K_{i\bar{j}l}-\frac{2}{3}K_{i\bar{j}}K_{\bar{l}})]\bar{D}^{\dot{\alpha}}\bar{\chi}^{\bar{j}}\bar{D}_{\dot{\alpha}}\bar{\chi}^{\bar{l}}-\bar{\phi}^{-1}\frac{D^{2}}{4}(e^{K/3}\phi^{2}D^{\bar{k}}W)\label{KGeqnforchi}\end{equation}

Let us now consider the case with one heavy superfield $H$ which
we will take to be canonically normalized. So the Kahler potential
becomes,\begin{equation}
K=\bar{H}H+K^{l}(L,\bar{L})\label{HLkahler}\end{equation}

where $L$ stands for the light fields. Also the superpotential is
taken to be \[
W=\frac{1}{2}MH^{2}+\tilde{W}(H,L)\]

where $M$is a large mass parameter.

Then from (\ref{KGeqnforchi}) we have for the heavy field 

\begin{eqnarray}
-4\bar{\phi}\square\bar{H} & =D^{2}(e^{K/3M_{p}^{2}}\phi^{2})D_{H}W+4Me^{2K/M_{p}^{2}}\phi^{2}\bar{\phi}^{2}(1+\frac{\bar{H}H}{M_{p}^{2}})D_{\bar{H}}\bar{W}\nonumber \\
+ & (1+\frac{\bar{HH}}{M_{p}^{2}})[e^{K/3M_{p}^{2}}\phi^{2}((\frac{2}{3}\frac{K_{l}}{M_{p}^{2}}D^{\alpha}\chi^{l}-\frac{2D^{\alpha}\phi}{\bar{\phi}})+\frac{\bar{H}D^{\alpha}H}{M_{p}^{2}}) & +2D^{\alpha}(e^{K/3M_{p}^{2}}\phi^{2})]D_{\alpha}H\nonumber \\
- & \frac{\bar{\phi}}{6M_{p}^{2}}D^{2}K_{\bar{l}}\bar{D}^{\dot{\alpha}}\bar{H}\bar{D}_{\dot{\alpha}}\bar{\chi}^{\bar{l}}+2D^{\alpha}(e^{K/3M_{p}^{2}}\phi^{2})D_{\alpha}D_{H}\tilde{W} & +e^{K/3M_{p}^{2}}\phi^{2}D^{2}D_{H}\tilde{W}\label{boxH}\end{eqnarray}

In the above $D_{H}W=\partial_{H}W+K_{H}W/M_{p}^{2}$ is the Kaehler
derivative of the superpotential with respect to the heavy field and
we have restored the dependence on the Planck mass. To integrate out
a heavy field $H$ we have to set $\square\bar{H}\rightarrow p^{2}\bar{H}\rightarrow0$
and the condition for that is that the right hand side of the above
equation is set to zero. Note that this condition reduces to (\ref{susyheavy})
in the global limit $M_{p}\rightarrow\infty$ and $\phi\rightarrow1$.
Here up to terms involving a factor of $D_{\alpha}H$ we have \[
D^{2}(e^{K/3M_{p}^{2}}\phi^{2})D_{H}W+4Me^{2K/M_{p}^{2}}\phi^{2}\bar{\phi}^{2}(1+\frac{\bar{H}H}{M_{p}^{2}})D_{\bar{H}}\bar{W}=-e^{K/3M_{p}^{2}}\phi^{2}D^{2}D_{H}\tilde{W}\]

As in the case of the global SUSY discussion one may expect that with
some restriction on the light field space (so that the right hand
side of the equation is small compared to $M$) the relevant condition
would be the natural generalization of (\ref{susyextrem})\begin{equation}
D_{H}W=\partial_{H}W+\frac{1}{M_{p}^{2}}W\partial_{H}K=0\label{heavykaehler}\end{equation}

Note that (by taking its spinor derivative) this condition implies
\[
W\bar{D}_{\dot{\alpha}}\bar{H}=0,\]
 so that the other $O(M)$ terms which all have a factor of $D_{\alpha}H$
also vanish (since the superpotential should not vanish at a generic
point). So (\ref{heavykaehler}) is certainly a sufficient condition
in the sense that it implies $\square H=0$ . 

However it is easy to see that the strict implementation of the condition
(\ref{heavykaehler}) leads to the conclusion that the light fermion
fields would have to be set to zero. Let us look at this in somewhat
more general terms than above.

We assume that the Kaehler potential is a sum of terms as in the string
theory examples discussed in \cite{deAlwis:2005tf}. In particular
if we call the heavy superfields $H^{I}$ and the light superfields
$L^{i}$ assume that\begin{equation}
K=K^{h}(H,\bar{H)}+K^{l}(L,\bar{L})\label{directsumK}\end{equation}
 Thus in the example in section 4 below, $H^{I}=z^{i}$and $L^{i}=S,T$.
Then the generalization of (\ref{heavykaehler}) becomes (note that
capital letters $I,J$ go over the heavy fields)\begin{equation}
K^{h\bar{I}J}D_{J}W=0\label{Heqn}\end{equation}

Using the non-degeneracy of the metric on the heavy fields this becomes,

\begin{equation}
\partial_{I}W+K_{I}^{h}W=0\label{FeqnforH}\end{equation}
 Taking the anti-chiral derivative of this equation and using the
chirality of the superpotential we get,

\begin{equation}
WK_{I\bar{J}}^{h}\bar{\mathcal{D}}_{\dot{\alpha}}\bar{H}^{\bar{J}}=0\label{antichiral}\end{equation}

implying $\bar{\mathcal{D}}_{\dot{\alpha}}\bar{H}^{\bar{J}}=0$ since
the metric is $K_{I\bar{J}}^{h}$ is non-degenerate and $W\ne0$ at
generic points.

Now let us assume that there is a solution $H=H(L,\bar{L})$ of equation
(\ref{FeqnforH}). Using the chirality of $H$ and $L$ we get by
differentiating this solution,\begin{equation}
\bar{\mathcal{D}}_{\dot{\alpha}}H=\frac{\partial H^{I}}{\partial L^{j}}\bar{\mathcal{D}}_{\dot{\alpha}}L^{j}+\frac{\partial H^{I}}{\partial\bar{L}^{\bar{j}}}\bar{\mathcal{D}}_{\dot{\alpha}}\bar{L}^{\bar{j}}=\frac{\partial H^{I}}{\partial\bar{L}^{\bar{j}}}\bar{\mathcal{D}}_{\dot{\alpha}}\bar{L}^{\bar{j}}=0.\label{eq:3}\end{equation}

Similarly from the result (see (\ref{antichiral})) that the chiral
derivative of $H$ also vanishes, \begin{equation}
\mathcal{D}_{\alpha}H^{I}=\frac{\partial H^{I}}{\partial L^{j}}\mathcal{D}_{\alpha}L^{j}+\frac{\partial H}{\partial\bar{L}^{\bar{j}}}\mathcal{D}_{\alpha}\bar{L}^{\bar{j}}=\frac{\partial H^{I}}{\partial L^{j}}\mathcal{D}_{\alpha}L^{j}=0\label{eq:4}\end{equation}

These two equations tell us that the chiral derivative of $L$ should
be zero - in other words the light fermions should also be set to
zero. This of course means that the light field theory is not supersymmetric!
However the problem is that the condition (\ref{heavykaehler}) or
its generalization is really too strong. It was obtained by ignoring
the $D_{\alpha}H$ terms as well as the $O(1/M)$ terms. Of course
as was pointed out earlier, these terms are zero if one imposes (\ref{heavykaehler})
so that one gets $\square H=0$, but the condition itself is not necessary.
The actual necessary condition which follows from (\ref{boxH}) is
a relation between the Kahler derivative terms and the fermionic superfield
terms. This condition would then express the bosonic component of
the heavy superfield in terms of the light bosonic field as well as
squares of light fermionic fields. However if we are interested only
in the scalar potential we do not need to keep these terms. Thus the
condition (\ref{heavykaehler}) can clearly be still used if one is
just interested in computing the potential for the light chiral scalars.

Unlike the case of rigid supersymmetry (\ref{heavykaehler}) is not
a holomorphic equation since the Kaehler derivative involves the real
function $K(\Phi,\bar{\Phi)}$. This means that the solution for the
heavy field will not in general be a holomorphic function of the light
fields and hence the light field theory in general will have a superpotential
that is just one (or a constant) and the whole effect of integrating
out the heavy field will be accounted for by changing the Kaehler
potential. In fact the original potential should be expressed in terms
of the Kaehler invariant function $G=\ln K+\ln|W|^{2}$before integrating
out the heavy fields. In a companion paper \cite{deAlwis:2005tf}
we show this explicitly in some examples that come from type IIB string
theory compactifications with fluxes. 

Now one might worry that the restriction on the range of the light
field essentially forces us back to the global case. This would indeed
be the case in an example such as (\ref{example}). Evaluating (\ref{heavykaehler})
in this case we have,

\[
D_{H}W=MH+\frac{1}{2}L^{2}+\bar{H}\frac{1}{2}(MH^{2}+HL^{2})=0\]

Solving this for $H$ we may compute the scalar potential using the
standard supergravity formula (expressed in terms of $G$) but now
the question is whether it is consistent to keep the supergravity
corrections given that the integrating out formula above, is valid
only for $|a|^{2}/M^{2}<<1$ as in the global case discussed earlier.
The point is that necessarily $M\le M_{p}$ so that supergravity corrections
which in this model are $O(|a|^{2}/M_{p}^{2})$ should also be ignored
for consistency of the approximation.

However very often in string theoretic examples (such as those with
flux compactifications) there is a constant in the superpotential
$W=W_{0}+...$ where the ellipses denote field dependent terms. In
these cases the supergravity corrections are indeed significant since
$D_{L}W=\partial_{L}W+K_{L}W/M_{p}^{2}\simeq\partial_{L}W+K_{L}W_{0}/M_{P}^{2}$
and the second term may even be of $(O(1)$ even thought light field
space is restricted to $|a|^{2}<<M^{2}$. So this does not necessarily
force us to the global limit since in many examples of interest in
string theory there would be a constant in the superpotential which
is generically of the order of the Planck/String scale. Thus the extra
piece in the Kaehler derivative (as compared to the ordinary derivative)
of the superpotential has to be kept. So we may use the condition
(\ref{Heqn}) with the understanding that it is to be used only for
the scalar components of the superfields for the purpose of calculating
the light scalar field potential, still remaining within the context
of supergravity.

\textbf{Acknowledgments: }

I'm very grateful to Martin Rocek for several useful suggestions.

\bibliographystyle{/usr/users/dealwis/apsrev}
\bibliography{myrefs}

\end{document}